\documentclass[pra,twocolumn,floatfix,superscriptaddress]{revtex4-1}

\usepackage{amsmath,amsfonts}
\usepackage{bbold}
\usepackage{graphicx}
\usepackage{dcolumn}
\usepackage{booktabs}
\usepackage{amsfonts}
\usepackage{color}
\usepackage[
breaklinks,
colorlinks,
linkcolor=blue,
citecolor=blue,
urlcolor=blue]{hyperref}

\newcommand{\tr}{\mbox{Tr}}

\begin{document}

\title{Efficient temperature-dependent Green's function methods for
realistic systems: using cubic spline interpolation to approximate
Matsubara Green's functions}

\author{Alexei A. Kananenka}
\email{akanane@umich.edu}
\author{Alicia Rae Welden}
\affiliation{Department of Chemistry, University of Michigan,
Ann Arbor, Michigan 48109, United States}
\author{Tran Nguyen Lan}
\altaffiliation{On leave from Ho Chi Minh City Institute of Physics, VAST, 
Ho Chi Minh City, Vietnam.}
\affiliation{Department of Chemistry, University of Michigan,
Ann Arbor, Michigan 48109, United States}
\affiliation{Department of Physics, University of Michigan,
Ann Arbor, Michigan 48109, United States}
\author{Emanuel Gull}
\affiliation{Department of Physics, University of Michigan,
Ann Arbor, Michigan 48109, United States}
\author{Dominika Zgid}
\affiliation{Department of Chemistry, University of Michigan,
Ann Arbor, Michigan 48109, United States}

\date{\today}


\begin{abstract}

The popular, stable, robust and computationally inexpensive cubic spline 
interpolation algorithm is adopted and used for finite temperature 
Green's function calculations of realistic systems. We demonstrate that with 
appropriate modifications the temperature dependence can be preserved while the 
Green's function grid size can be reduced by about two orders of magnitude by replacing 
the standard Matsubara frequency grid with a sparser grid and a set of interpolation coefficients. 
We benchmarked the accuracy of our algorithm as a function of a single parameter 
sensitive to the shape of the Green's function. Through numerous examples, we confirmed 
that our algorithm can be utilized in a systematically improvable, controlled, and 
black-box manner and highly accurate one- and two-body energies and one-particle 
density matrices can be obtained using only around 5\% of the original grid points. 
Additionally, we established that to improve accuracy by an order of magnitude, the number 
of grid points needs to be doubled, whereas for the Matsubara frequency grid 
an order of magnitude more grid points must be used. This suggests that 
realistic calculations with large basis sets that were previously out of reach 
because they required enormous grid sizes may now become feasible.

\end{abstract}

\maketitle


\section{Introduction}

Finite-temperature Green's function calculations have a long history in 
condensed matter physics~\cite{Fetter,Abrisikov_Gorkov_Dziloshinsky,Mattuck,
stefanucci2013nonequilibrium}. Most commonly the finite temperature formalism 
is employed to study low-energy effective models, such as the Hubbard~\cite{Hubbard:prsla/276/238} 
model, either by numerical or analytical 
means~\cite{Imada:rmp/70/1039,Gull:rmp/83/349,Georges:rmp/68/13,Kotliar:rmp/78/865,Scalapino:rmp/84/1383}. 
Much less is known about employing finite temperature Green's functions 
for realistic systems while maintaining ``chemical accuracy'' of 1 kcal/mol.
In quantum chemistry, or in any realistic calculations beyond model systems, 
the eigenvalue spread of realistic Hamiltonians is very broad; thus, the 
finite temperature Green's function formalism, when applied to realistic problems, 
has to describe energy levels far outside the window described by 
model Hamiltonians. This makes the realistic calculations much more challenging than the 
model ones. In practical calculations, temperature-dependent Green's functions are 
expressed as functions of imaginary time or imaginary frequency (Matsubara Green's 
function) that are mutually connected by a Fourier Transform. Both imaginary time
and Matsubara Green's functions are discretized on imaginary time and imaginary frequency 
grid, yielding $G(1:n,1:n,1:N_{\tau})$ and $G(1:n,1:n,1:N_{\omega})$ matrices, where $n$, 
$N_{\tau}$, and $N_{\omega}$ are the number of orbitals, imaginary time, and imaginary frequency grid 
points, respectively. The size of these grids depends on the energy spread of a system and the 
temperature-dependent grid spacing. Since realistic Hamiltonians have a wide spread of orbital 
energies, both grids need to contain hundreds of thousands of grid points to reach high numerical 
accuracy at low temperatures that result in small grid spacing.
Even though in a parallel calculation each grid point can be processed independently, 
computation requirements are still very high and  both the time and memory necessary 
to handle Green's function operations grow steeply. 
Thus, for accurate and 
affordable realistic calculations, it is highly desirable to find compact representations 
of both imaginary time and Matsubara Green's functions.

Recently, Boehnke et al.~\cite{Boehnke:prb/84/075145} employed orthogonal polynomial 
representation of Green's functions to compactly express Hubbard Green's functions.
Using this approach for realistic systems, we have shown that 
very accurate results can be obtained exploiting only a fraction of the original 
imaginary time grid points necessary to illustrate the energy spread of the realistic 
Hamiltonian.~\cite{Kananenka:jctc/12/564}
In practical calculations, since many Green's functions manipulations are easier
in frequency space it is important to have a compact representation of the 
Matsubara Green's function. In this paper, we focus on finding a representation that 
will result in using compact Matsubara frequency grids for realistic problems.

The regular imaginary frequency Matsubara grid is equidistant and the grid spacing is 
directly related to the physical temperature. Let us note, however, that while the grid 
spacing for low frequencies is essential to illustrate the physical temperature, for 
larger frequencies the Green's function is a slowly and smoothly changing function of 
frequency. Consequently, it should be possible to keep the original spacing for few 
frequency points near zero and have a prescription to systematically evaluate more points 
for higher frequencies with larger than the near-zero spacing without any loss of accuracy. 

Since the Matsubara Green's function or self-energy is smoothly and slowly changing 
between grid points, numerical interpolation is especially suitable to accurately describe it. 
Linear interpolation is the simplest choice but it lacks smoothness. While a polynomial of higher 
degree may be used to interpolate and ensures smoothness, this type of interpolation may result in large 
Runge oscillations between the data points. Consequently, we decided to employ spline interpolation 
using cubic polynomials over a polynomial interpolation since it will result in a procedure with 
much smaller interpolation errors, greater stability and low computational cost. 

In chemical physics, cubic spline interpolation has been applied as a basis to solve a complex 
differential and integral Schr\"odinger~\cite{Shore:jcp/58/3855,Shore:jpb/6/1923,ODERO:ijmpc/12/1093}, 
Dirac~\cite{Johnson:pra/37/307}, and Sham--Schl\"uter equations~\cite{Hellgren:prb/76/075107}, 
Thomas--Fermi model~\cite{Raynor:cp/66/409}, in calculations of vibrational and rotational 
spectra~\cite{Cremaschi:mp/41/759} and many other 
cases~\cite{Williamson:prl/87/246406,Gilbert:cpl/29/569,Brosolo:cp/159/185,Jiang:jcp/95/4044,Shore:jpb/8/2023,
Gazquez:jcp/67/1887,Carroll:jcp/71/4142,Hutchinson:jcp/85/7087,Jiang:jcp/87/6973,Decleva:ijqc/56/27,Bachau:rpp/64/1815,
Thulstrup:ijqc/9/789,Darcy:jcp/143/074311}. Since the derivatives of third and higher order 
polynomials are discontinuous, cubic spline interpolation is limited to applications that are 
not sensitive to the smoothness of derivatives higher than second order. 
Applications of cubic spline interpolation algorithm in Green's function theory are known in the 
context of dynamical mean-field theory (DMFT)~\cite{blumer:phd,Georges:rmp/68/13,Maier:rmp/77/1027}.

Finally, cubic spline interpolation algorithm is popular because it is 
very easy to implement and use. Several libraries provide built-in functions for cubic spline 
interpolation. For example, FORTRAN provides both procedural and object-oriented interfaces for the 
FITPACK library~\cite{dierckx1995curve}.

This paper is organized as follows. In section~\ref{theory}, we shortly review the background 
necessary to understand finite temperature Green's functions and our motivation behind applying 
them to realistic calculations. Additionally, we focus on illustrating the difficulties of extending 
the Green's function approach to large basis sets used in chemistry. In section~\ref{comp_details}, 
we describe the spline interpolation procedure that we use for realistic systems and its implementation 
in the second-order Green's function theory. We list and discuss the numerical results of our algorithm 
as applied to realistic atomic and molecular calculations in section~\ref{results}. Finally, we 
present conclusions in section~\ref{conclusions}.

\section{Theory}
\label{theory}

In this section, we briefly review some  aspects of fermionic Green's function theory
relevant to this work. For a more detailed introduction to Green's function
theory readers are suggested to consult textbooks on the subject, see e. g. 
ref~\citenum{Fetter,Jishi,stefanucci2013nonequilibrium}. 

First, let us note that Green's functions can be expressed in the real or imaginary frequency domain.
In general a real time or real frequency one-body Green's function is a function used to 
describe spectral properties such as ionization potentials, electron affinities, or the single-particle
spectral function. Methods such as the random phase approximation (RPA) 
and GW usually express the Green's function using real frequencies to obtain spectra at zero 
temperature~\cite{doi:10.1021/ct5001268,Govoni:jctc/11/2680}.
While, in general, the real frequency Green's function is an exponentially decaying function of frequency, 
it is very difficult to employ it in iterative methods such as DMFT
or other embedding methods such as self-energy embedding 
theory (SEET)~\cite{Kananenka:prb/91/121111,Lan:jcp/143/241102} since iterating usually requires pole 
shifting algorithms~\cite{Lu:prb/90/085102,Neck:jcp/115/15,Peirs:jcp/117/4095}.

The imaginary frequency Matsubara Green's function $\mathbf{G}(i\omega_n)$ is used to describe single-particle
properties of a statistical ensemble where many excited states (besides the ground state) are potentially 
accessible at a given finite temperature. While not commonly employed in molecular quantum 
chemistry calculations, such Green's functions are desirable for materials science calculations where a small 
electronic band gap allows multiple electronic states to be populated even at low temperatures. 
The Matsubara Green's function is expressed on the  imaginary grid
$i\omega_n=(2n+1)\pi/\beta$~\cite{Matsubara:ptp/14/351}, 
where $n = 0,1,2,...$, $\beta=1/(k_\text{B}T)$ is the inverse temperature, and $k_{\rm B}$ is the Boltzmann constant. 
Note that the  $2\frac{\pi}{\beta}$ spacing of the grid is set by the physical temperature $T$.
Using such a grid, the Matsubara Green's function is then defined as 
 \begin{equation}
\mathbf{G}(i\omega_n) = \left[ (i\omega_n + \mu)\mathbf{S} - \mathbf{F} - 
\mathbf{\Sigma}(i\omega_n) \right]^{-1},
\end{equation}
where $\mathbf{S}$ and $\mathbf{F}$ are the overlap and Fock matrices correspondingly, and 
$\mu$ is the chemical potential chosen such that a proper number of electrons is present in the system.
The self-energy $\mathbf{\Sigma}(i\omega_n)$ is a correction to the non-interacting Green's function 
$\mathbf{G}_0(i\omega_n) = \left[ (i\omega_n + \mu)\mathbf{S} - \mathbf{F} \right]^{-1}$ describing
static and dynamical many-body correlation effects at the single-particle level.

Both real and imaginary parts of the Green's function on a Matsubara grid are smooth and converge to zero in 
the limit of large frequencies. In this high-frequency limit, $i\omega_n \to \infty$, the Matsubara Green's 
function can be expressed as a series
\begin{equation}
\mathbf{G}(i\omega_n) = \frac{\mathbf{G}_1}{i\omega_n} + 
\frac{\mathbf{G}_2}{\left(i\omega_n \right)^2} + 
\frac{\mathbf{G}_3}{\left(i\omega_n \right)^3} 
+ \mathcal{O}\left( \frac{1}{\left (i\omega_n \right)^4}\right),  \label{hfe}
\end{equation}
with the expansion coefficients given by
\begin{equation}
\left[ G_{k} \right]_{ij} = (-1)^{(k-1)} \langle \Psi | \{ \left[ \hat{H},\hat{c}_i 
\right]_{k},\hat{c}_j^\dagger \} | \Psi \rangle,	\label{hfe2}
\end{equation}
where $\hat{H}$ is the full many-body Hamiltonian of the system
\begin{equation}
\hat{H} = \sum_{ij}^n h_{ij} \hat{c}_i^\dagger \hat{c}_j + 
\frac{1}{2}\sum_{ijkl}^n v_{ijkl} \hat{c}_i^\dagger \hat{c}_k^\dagger \hat{c}_l \hat{c}_j,
\end{equation}
where $\hat{c}_i$ ($\hat{c}^\dagger_i$) is the electron annihilation (creation) operator from orbital $i$,
$h_{ij}$ is the core-Hamiltonian matrix and $v_{ijkl}$ are two-electron integrals defined as
\begin{equation}
v_{ijkl} = \int \int d{\bf r}_1 d{\bf r}_2 \phi^{*}_{i}({\bf r}_1)\phi_{j}({\bf r}_1)\frac{1}{r_{12}}
\phi^{*}_{k}({\bf r}_2)\phi_{l}({\bf r}_2).
\end{equation}
$|\Psi\rangle$ is the Heisenberg ground state of the system.
It was shown in ref~\citenum{Rusakov:jcp/141/194105} that the coefficients of high-frequency 
expansion in a non-orthogonal orbital basis for Hamiltonians with full Coulomb interaction are given by
\begin{align}
\mathbf{G}_1 = & \mathbf{S}^{-1}, \\ 
\mathbf{G}_2 = & \mathbf{S}^{-1}\left( \mathbf{F} - \mu \mathbf{S} \right)\mathbf{S}^{-1}.
\label{eq:g_coeffs}
\end{align}

\begin{figure}[htbp]
\begin{center}
\includegraphics[width=0.49\textwidth]{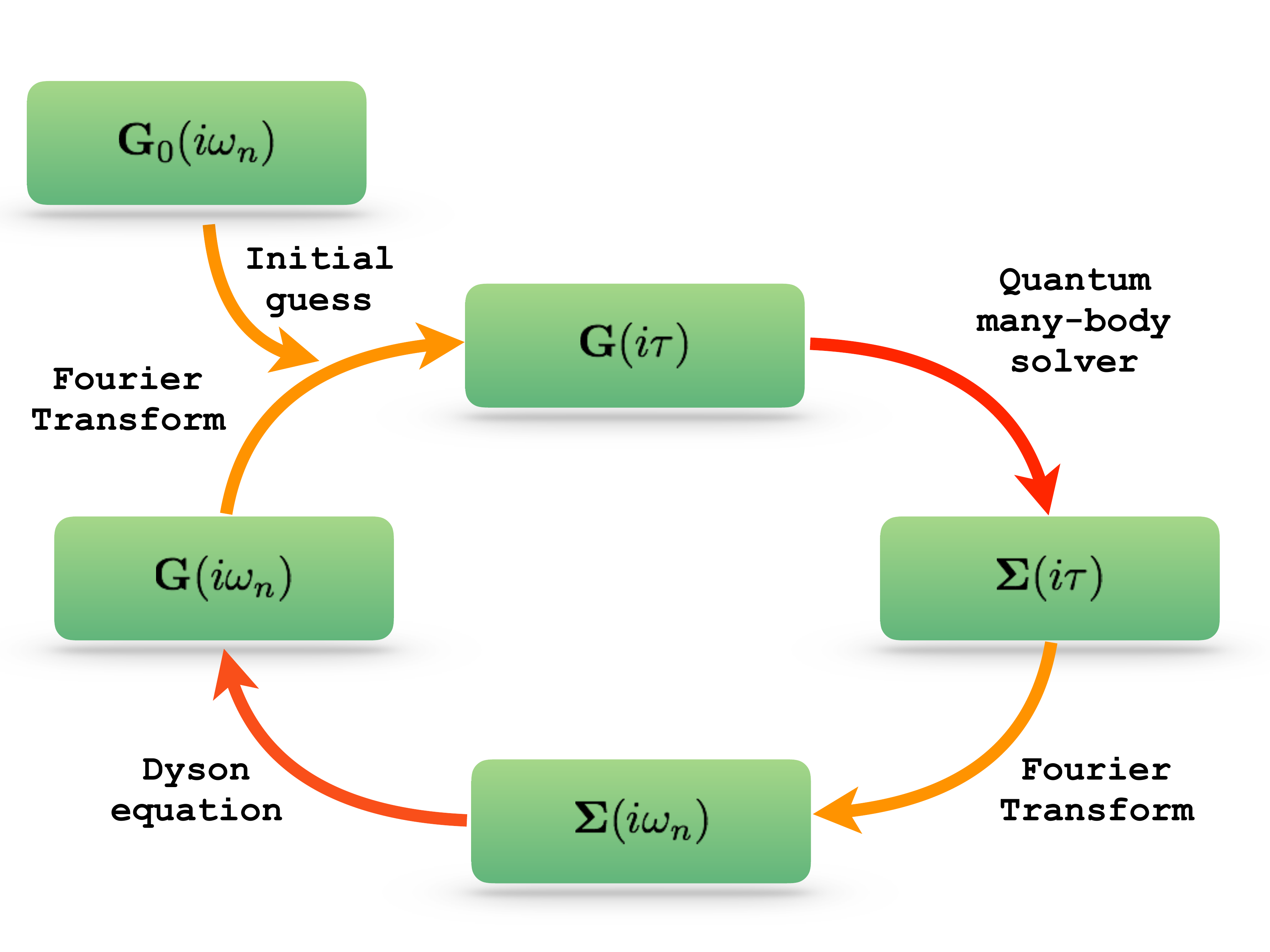}
\caption{A typical self-consistent Green's function calculation consists of the following steps. Generating an initial guess
is followed by the Fourier transform from Matsubara domain to imaginary time domain. Then a quantum many-body
problem is solved using many-body solver. This step is usually done in the imaginary time domain. The inverse
Fourier transform back to Matsubara domain and solution of the Dyson equation conclude the iteration and
the next iteration starts with updated Green's function. The most computationally expensive steps are the 
solution of quantum many-body problem and solution of the Dyson equation.}
\label{fig:gf_flow}
\end{center}
\end{figure}

In a typical calculation, see Fig.~\ref{fig:gf_flow}, the self-energy 
$\mathbf{\Sigma}$ is evaluated either on the Matsubara frequency or imaginary time grid by a variety of solvers 
ranging from quantum Monte Carlo methods~\cite{Gull:rmp/83/349,Rubtsov:prb/72/035122, Werner:prl/97/076405,Werner:prb/74/155107} 
to perturbative~\cite{Georges:prb/45/6479,Kajueter:prl/77/131,Dahlen:jcp/122/164102,Phillips:jcp/140/241101,
Hedin:pr/139/A796,Aryasetiawan:rpp/61/237} and configuration interaction type of 
methods~\cite{Zgid:prb/86/165128,Caffarel:prl/72/1545}. In the next step, a Green's function is calculated by means of 
the Dyson equation
\begin{equation} 
\mathbf{G}(i\omega_n)^{-1} = \mathbf{G}_0(i\omega_n)^{-1} - \mathbf{\Sigma}(i\omega_n).
\label{eq:dyson}
\end{equation}

Using the correlated Green's function and self-energy, one evaluates quantities of interest such as the
one-particle density matrix  
\begin{equation}
\mathbf{P} = \frac{2}{\beta}\sum_n e^{i\omega_n 0^+}\mathbf{G}(i\omega_n),
\label{eq:gamma}
\end{equation}
and using it the correlated  one-body energy
\begin{equation}
E_\text{1b} = \frac{1}{2}\tr \left[ \left(\mathbf{h} + \mathbf{F} \right) \mathbf{P} \right],
\label{eq:e1b}
\end{equation}

Different prescriptions can be used to compute the two-body correlation energy $E_\text{2b}$. 
The Galitskii--Migdal formula~\cite{Galitskii:jetp/34/139} is used to evaluate the internal energy 
\begin{equation}\label{GM_enrg}
E^\text{GM}_\text{2b} = \dfrac{2}{\beta} \sum_n^{N_\omega} \tr 
\left[\mathbf{G}(i\omega_n)\mathbf{\Sigma}(i\omega_n) \right],
\end{equation}
where $N_\omega$ is the total number of imaginary frequencies.  
Similarly, the Luttinger--Ward~\cite{Luttinger:pr/118/1417} functional $\Phi$ consisting of 
irreducible energy diagrams of the self-energy can be used to calculate the grand-canonical potential 
\begin{equation}\label{LW_enrg}
\Omega[\mathbf{G}] = \tr \left[ \ln \left( \mathbf{\Sigma} - \mathbf{G}_0^{-1} \right) + 
\mathbf{\Sigma} \mathbf{G} \right] - \Phi[\mathbf{G}],
\end{equation}
which at low temperatures reduces to $\Omega = E - \mu N$, where $E = E_\text{1b} + E_\text{2b}$ is the 
internal energy and $N$ is the total number of electrons in the system. For higher temperatures, 
$\Omega = E - TS - \mu N$ 
can be used to find the free energy and thermodynamic properties of the system under study.
For self-consistent calculations at low temperature the stationary value of $\Omega$
corresponds to the energy obtained from Galitskii--Migdal formula up to a shift of $\mu N$.

In practical calculations, a finite number of imaginary frequencies $N_\omega$ is 
used to span the Matsubara Green's function. Insufficient size of the Matsubara frequency grid
leads to errors in the one- and two-body energy as well as the one-particle density matrix.

\begin{figure*}[htbp]
\begin{center}
\includegraphics{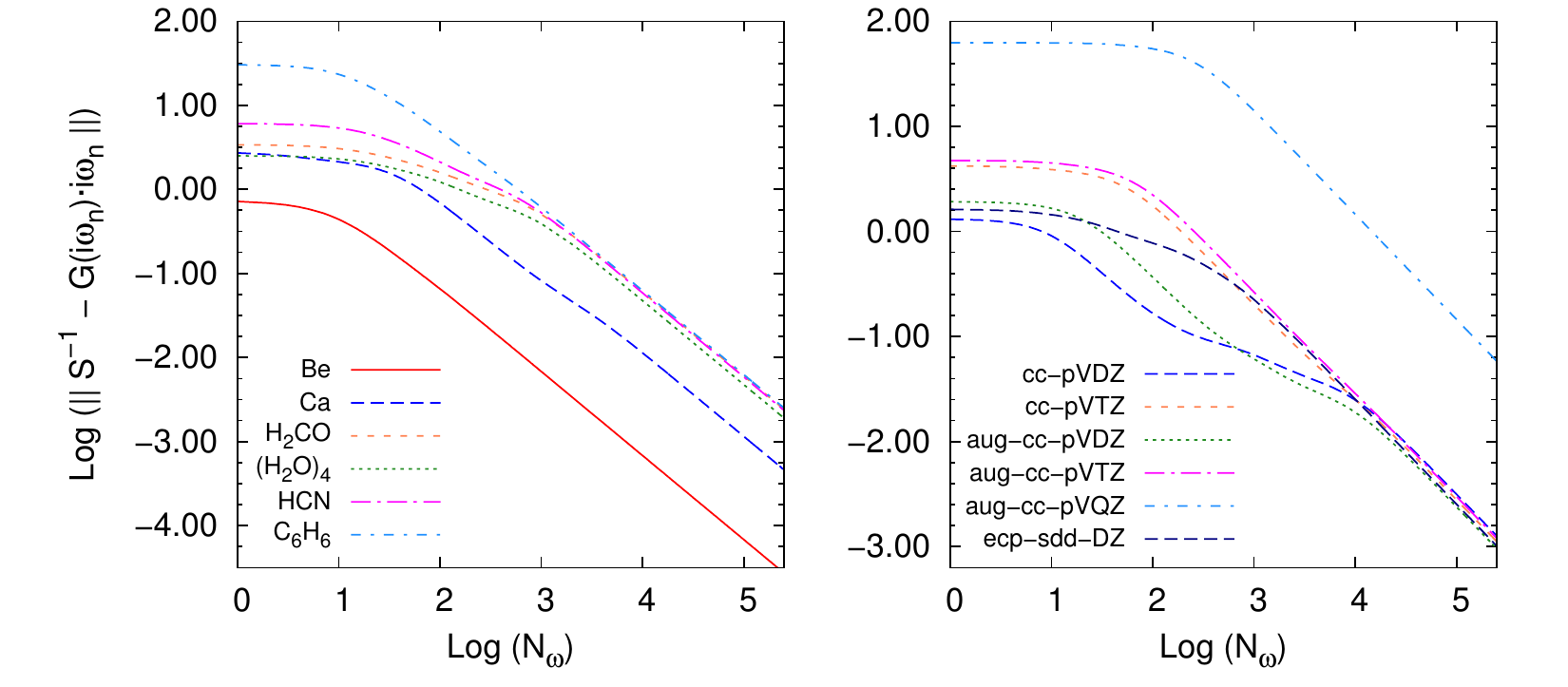}
\caption{Left panel: The convergence of  $ || \mathbf{D}_{\mathbf{G}_1}||_\text{F}$ for the Hartree--Fock Green's function 
 as a function of the number of Matsubara frequencies. We used aug-cc-pVDZ basis set for Be, 
cc-pVDZ~\cite{Koput:jpca/106/9595} basis set for Ca and TZ(Dunning)~\cite{Dunning:jcp/55/716} basis set 
for molecules. Right panel: The effect of the basis set on the convergence of  $ || \mathbf{D}_{\mathbf{G}_1}||_\text{F}$ 
for the Kr atom, $\beta$=100 [1/a.u]. Log denotes the base 10 logarithm.}
\label{fig:s_norm}
\end{center}
\end{figure*}

A practical way to decide how many frequencies 
have to be used to represent the Matsubara Green's function before applying the high-frequency expansion 
can be based on measuring the distance between the inverse of the overlap matrix $\mathbf{S}^{-1}$ and 
the numerically evaluated coefficient of high-frequency expansion of Green's function 
$\mathbf{G}_1=\mathbf{G}(i\omega_n)\cdot i\omega_n$ . In the limit of infinite number of Matsubara frequencies 
$ \mathbf{D_{\mathbf{G}_1}}=\lim_{N_\omega \to \infty}\left( \mathbf{G}(i\omega_n)\cdot i\omega_n - \mathbf{S}^{-1} \right) \to \mathbb{0}$. 
A finite number of Matsubara frequencies always results in an error. To illustrate the magnitude of this error,
we plot in the left panel of Fig.~\ref{fig:s_norm} the Frobenius norm 
$|| \mathbf{D}_{\mathbf{G}_1}||_\text{F} \equiv \sqrt{ \sum_i^m \sum_j^n | (D_{G_1})_{ij} |^2}$ of 
$\mathbf{D_{\mathbf{G}_1}}$
as a function of the number of Matsubara frequencies $N_\omega$ for seven realistic atomic and molecular systems. 
After an initial plateau, where no improvement is seen, the Frobenius norm starts to decay linearly in the 
logarithmic plot. Consequently, to reduce the error in the Frobenius norm by an order of magnitude,
an order of magnitude more Matsubara frequencies is necessary.
The right panel of Fig.~\ref{fig:s_norm} shows the convergence of $|| \mathbf{D}_{\mathbf{G}_1}||_\text{F}$ as a function of 
a basis set for the Kr atom. Generally, employing larger basis sets or adding diffuse functions requires
an increase of the number of Matsubara frequencies. This is due to the fact that the spread of Hamiltonian 
eigenvalues increases as larger basis sets are used. Thus, the fastest decay of $|| \mathbf{D}_{\mathbf{G}_1}||_\text{F}$ 
is observed in cc-pVDZ~\cite{Wilson:jcp/110/7667} basis with 27 basis function and the slowest in the 
aug-cc-pVQZ~\cite{Wilson:jcp/110/7667} with 97 functions.
Note that even when a basis with a pseudopotential that has only basis functions describing the 
valence orbitals is employed, to be in the linear regime requires more than 10,000 frequencies.

During a calculation, the Matsubara Green's function may need to be stored in memory. 
Storing a single Green's function requires $\mathcal{O}(N_\omega n^2)$ complex double
precision numbers, where $n$ is the number of orbitals in a basis set. Thus, for the large orbital bases 
and large number of frequencies necessary to reach quantum chemical quantitative accuracy, the 
required memory becomes a significant bottleneck.
Even if the memory bottleneck is avoided and the Green's function is evaluated one the fly, 
when necessary, the computational complexity of all operations involving the Green's functions grows 
rapidly with the number of Matsubara frequencies. 
Examples of such operations include solving the Dyson equation~\ref{eq:dyson},
which requires $\mathcal{O}(N_\omega n^3)$ evaluations.
Even if all these operations can be made parallel over the frequency index, having to take into
account this large number of Matsubara frequencies can significantly slow down a computation.

\begin{figure*}[htbp]
\begin{center}
\includegraphics[width=0.99\textwidth]{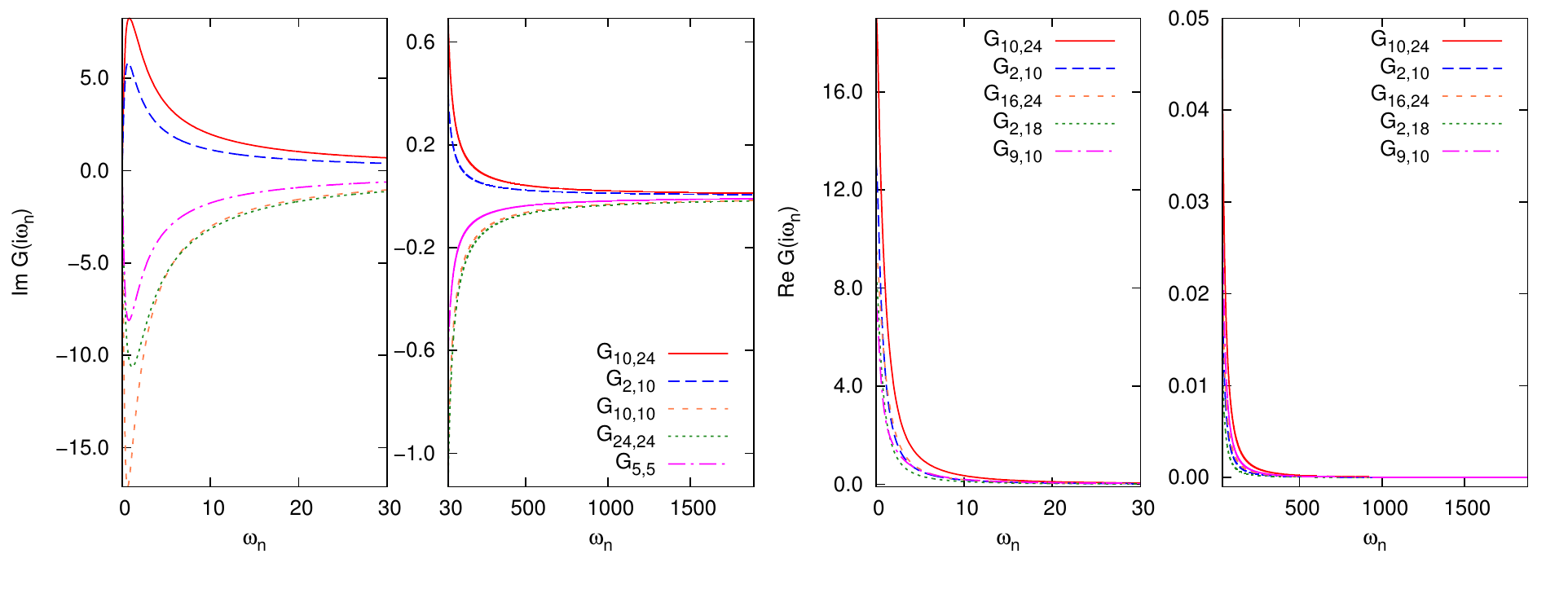}
\caption{The five largest elements of $\text{Im}\mathbf{G}(i\omega_n)$ 
(left two plots) and the five largest matrix elements of $\text{Re}\mathbf{G}(i\omega_n)$ (right two plots)
for H$_2$CO molecule calculated 
with the second-order Green's function perturbation theory GF2 with TZ(Dunning) basis set and $\beta$=100 [1/a.u]. 
Note that the frequency axis of the $\text{Im}\mathbf{G}(i\omega_n)$ and $\text{Re}\mathbf{G}(i\omega_n)$ is discontinuous  
to show that the most rapid change of the Green's function happens in the low-frequency range while for the 
remaining frequencies the Green's function converges slowly to the high-frequency limit.}
\label{fig:gf}
\end{center}
\end{figure*}

To see why the frequency grid requirements are so demanding when chemical accuracy is desired, 
it is instructive to look at Matsubara Green's functions evaluated in large basis sets. In the
left panel of Fig.~\ref{fig:gf}, we plotted several of the largest elements of the imaginary part of 
the Matsubara Green's function for the H$_2$CO molecule calculated using the second-order Green's 
function theory (GF2)~\cite{Holleboom:jcp/93/5826,Dahlen:jcp/122/164102,Phillips:jcp/140/241101} 
with TZ(Dunning) basis set. In the right panel of Fig.~\ref{fig:gf}, we 
plotted several largest matrix elements of the real part of the Matsubara Green's 
function for the H$_2$CO molecule.
First, let us note that different elements of $[G(i\omega_n)]_{ij}$ decay differently. 
Secondly, as expected,
the most rapid change in both real and imaginary parts of the Green's function occurs 
in the low-frequency range. Lastly, the slow convergence of the Green's function to its 
high-frequency limit is responsible for the large Matsubara grid when low temperatures are used. 
Similar observations can be made about the convergence of sums involving Matsubara 
frequencies~\cite{Nieto:cpc/92/54,Espinosa:mc/79/1709}.

Motivated by these observations, in the next section we will compute the Matsubara Green's function 
for only a few low-frequency points to preserve the temperature dependence and have 
various interpolation techniques to approximate values of the Green's function for higher frequencies. 

\subsection{Spline interpolation}
\label{spline}

We adopted the commonly used cubic spline interpolation technique and modified it appropriately for the 
efficient use in Green's function calculations. For the description of a standard cubic spline interpolation,  
readers are referred to textbooks on numerical methods, e.g. ref~\citenum{numrecC,Bradie}.

We consider a subset of Matsubara frequencies $\mathcal{S} = \{i\omega_{n} | n \in [0,N_\omega] \}$ chosen from original
equidistant Matsubara grid
and Green's function $G_{ij}(i\omega_n)$ evaluated on this subset. We focus on a particular frequency interval 
$\left[\omega_l,\omega_{l+1}\right], \omega_l,\omega_{l+1} \in \mathcal{S}$. 
We define a local polynomial between interval endpoints as a modified Legendre interpolation formula
\begin{equation}
\tilde{G}(i\omega_n) = a G(i\omega_l) + b G(i\omega_{l+1}) + c G^{''}(i\omega_l) + d G^{''}(i\omega_{l+1}),
\label{eq:gtilde}
\end{equation}
where $n \in [l,l+1]$, $G^{''}(i\omega_l)$ is the second derivative of Green's function at the point $i\omega_l$, and $a,b,c,d$
are interpolation coefficients determined as 

\begin{align}
a = & \frac{i\omega_{l+1} - i\omega}{i\omega_{l+1} - i\omega_l}, \\
b = & \frac{i\omega - i\omega_l}{i\omega_{l+1} - i\omega_l}, \\
c = & \frac{1}{6}\left(a^3 - a \right)\left(i\omega_{l+1} - i\omega_l\right)^2, \\
d = & \frac{1}{6}\left(b^3 - b \right)\left(i\omega_{l+1} - i\omega_l\right)^2. 
\end{align}

Polynomials $\tilde{G}(i\omega_n)$ from all intervals $[i\omega_l,i\omega_m] \subset [i\pi/\beta,i\omega_N]$
can be combined and used as an approximation
to the Green's function $G(i\omega_n)$. Equation~\ref{eq:gtilde} generates continuous second derivatives
both within the interval and at its boundaries, thus making a smooth transition between intervals, but 
approximating the third derivatives by a constant. 
Note, that a simple linear interpolation is not a good choice here because the second derivative is 
undefined at the boundaries of the intervals and is zero inside. A construction of a local polynomial (eq~\ref{eq:gtilde})
requires the knowledge of second derivatives for every given point $i\omega_l$.  Analytical second derivatives 
are not available for a correlated Matsubara Green's function and numerical derivatives must be used. 
The second derivatives can be approximated, for example, using central difference formula
\begin{equation}
G^{''}(i\omega_l) = \frac{G(i\omega_l + \zeta) - 2G(i\omega_l) - G(i\omega_l - \zeta)}{2\zeta},
\end{equation}
where $\zeta$ is a small increment. An application of the central difference formula for every interval results in the following 
$M-2$ equations for second derivatives

\begin{align}
& \frac{i\omega_l - i\omega_{l-1}}{6}G^{''}_{l-1} + \frac{i\omega_{l+1} - i\omega_{l-1}}{3}G^{''}_l + 
\frac{i\omega_{l+1} - i\omega_l}{6}G^{''}_{l+1} =\label{eq:sys1}  \\ \nonumber 
 = & \frac{G^{''}(i\omega_{l+1}) - G^{''}(i\omega_l)}{i\omega_{l+1} - i\omega_l} - 
\frac{G^{''}(i\omega_l) - G^{''}(i\omega_{l-1})}{i\omega_l - i\omega_{l-1}}.	 
\end{align}

Since there are only $M-2$ equations for $M$ unknowns to find a unique solution two more
equations have to be provided. In this work we chose to simply set second derivatives at global
boundaries to zero $G^{''}(i\omega_{n=0})=0$ and  $G^{''}(i\omega_{n=N_\omega})=0$. In the
numerical analysis literature, it is known as a natural spline~\cite{numrecC}.
Now, the resulting $M$ equations can be written as a matrix equation and finding second derivatives 
amounts to solving a system of linear equations. Since the coefficient matrix is tridiagonal
there is a unique solution that can be obtained in $\mathcal{O}(M)$ operations using a sparse
linear solver.

Once the second derivatives are known and coefficients $a,\dots,d$ are calculated, the Green's function can be 
reconstructed using eq~\ref{eq:gtilde} at any requested frequency point. Our algorithm consists of the following 
basic steps:

\begin{enumerate}

\item We begin by choosing a small number of grid points explicitly and forming a small preliminary 
grid that usually does not exceed a few hundred points. Most of these points are located near zero frequency and 
preserve the natural Matsubara spacing to encode information about the inverse temperature $\beta$. It 
is not particularly important how the points are chosen further away from zero frequency because later, 
iteratively, more points are added when necessary. However, to keep the number of operations small it is 
recommended to take advantage of the shape of Green's function and create a denser grid in the low-frequency 
region and a sparser grid everywhere else. 

\item At these preliminary grid points we evaluate the Green's function $G(i\omega)$.

\item \label{main} 
For every pair of labels $ij$ of the Green's function $[G(i\omega)]$, we solve a system of 
equations~\ref{eq:sys1} for second derivatives in every interval between adjacent grid points
and use it to infer the magnitude of change of the Green's function.

\item Since we use cubic polynomials the forth derivative $|G^{IV}|$ must vanish.
We calculate and compare the absolute value of $|G^{IV}|$ to the predetermined 
desired threshold value $\delta$. If $|G^{IV}| < \delta$ then the Green's function 
does not change on the 
$\left[i\omega_l,i\omega_{l+1}\right]$ interval appreciably and no more frequency 
points should be added, otherwise a midpoint
$i\omega_{(2l+1)/2}$ is inserted.

\item We evaluate and store the Green's function at the midpoint frequency $G(i\omega_{(2l+1)/2})$.

\item We repeat step~\ref{main} until $|G^{IV}|$ indicate that the Green's function does not change 
anymore on every interval for all the $ij$ labels.

\end{enumerate}

Depending on the value of $\delta$ a different number of imaginary frequency points are selected and hence the 
accuracy of the spline can be systematically improved by decreasing $\delta$.

Since the Matsubara Green's function is a complex quantity the cubic spline interpolation algorithm can be applied
to either real or imaginary part of it. The real part of the Green's function contributes to the density 
matrix and the one-body energy. The imaginary part of the Green's function influences the two-body energy. 
We observed that an insufficient grid causes the largest error in the density matrix and consequently the 
one-body energy. For this reason, we decided to apply the cubic spline interpolation algorithm to the real 
part of Green's function to minimize $\delta$ and we use the resulting grid to evaluate both the real and 
imaginary part of the Green's function.

\section{Computational details}
\label{comp_details}

The algorithm for creating a Green's function spline introduced above is suitable for calculating 
Green's functions or self-energies in a systematic manner and improving its accuracy as a function of 
the spline accuracy $\delta$ and the grid size. In this paper, we tested this algorithm on a series of Green's 
functions coming from GF2 calculations. GF2 is a perturbative many-body 
Green's function method that has many attractive properties. It is as accurate as M\o ller-Plesset 
perturbation theory (MP2)~\cite{Moller:pr/46/618} for weakly correlated systems but, unlike many 
methods suitable for weakly correlated systems such as MP2 or CCSD~\cite{Cizek:acp/14/35}, it is 
reasonably well behaved for 
strongly correlated systems~\cite{Phillips:jcp/140/241101}. GF2 has both small fractional charge and fractional spin 
errors~\cite{Phillips:jcp/142/194108}, affordable computational scaling $\mathcal{O}(N_\tau n^5)$ and 
can be carried out self-consistently, making it reference independent.
The self-consistency guarantees that the Luttinger--Ward functional constructed from the converged GF2 
Green's function and GF2 self-energy and the total energy is stationary with respect to the Green's function. 
Therefore, at convergence different ways of calculating correlation energy agree within numerical precision. 
This is a significant advantage because it means that one is free to choose the simplest way of evaluating the 
correlation energy e. g. using Galitskii--Migdal formula rather than Luttinger--Ward functional.

In GF2, the imaginary time self-energy $\mathbf{\Sigma}(i\tau)$ is calculated using an imaginary time Green's 
function $\mathbf{G}(i\tau)$ according to
\begin{align}
\Sigma_{ij}(i\tau) & =  -\sum_{klmnpq} G_{kl}(i\tau) G_{mn}(i\tau)
G_{pq}(-i\tau) \times \nonumber \\ 
& \times v_{ikmq} \left( 2v_{ljpn} - v_{pjln} \right ),
\label{eq:gf2}
\end{align}

GF2 calculations proceed as shown in Fig.~\ref{fig:gf_flow}
and operate in both imaginary time and Matsubara domains. This choice simplifies the numerical evaluation of the self-energy
and the solution of the Dyson equation, for details see ref~\citenum{Fetter,Mattuck}. A broad variety of complex numerical 
algorithms and procedures, involved in the GF2 calculation, require handling Matsubara grids such as the calculation of the
Matsubara Green's functions, fast Fourier transform from the Matsubara frequencies to imaginary time and back, 
solution of the Dyson equation and the evaluation of sums over Matsubara frequencies in the Galitskii--Migdal energy 
calculation. This makes GF2 an ideal candidate for testing our algorithm. The details of the GF2 algorithm can be
found in ref~\citenum{Phillips:jcp/140/241101}.
To accelerate calculations of the  imaginary time self-energy a Legendre polynomial basis was used as
described in ref~\citenum{Kananenka:jctc/12/564}. 

The reference data involving full frequency grid was obtained by performing self-consistent GF2 calculations for 
several atoms and simple molecules with Matsubara frequency grids large enough to achieve convergence in total energy to 10 
$\mu$E$_h$ and in the total number of electrons to $10^{-5}$. From the converged GF2 Green's functions and self-energies we 
calculated the reference one-body density matrix using eq~\ref{eq:gamma}, the total number of electrons, the one-body energy 
using eq~\ref{eq:e1b}, and the Galitskii--Migdal and Luttinger--Ward energies using eq~\ref{GM_enrg}-\ref{LW_enrg}
respectively. 
 
\section{Results and discussion}
\label{results}

In this section, we benchmark realistic atomic and molecular GF2 calculations using the cubic spline interpolation algorithm
described above. 
Our test set is comprised of 3 closed-shell atoms: Be, Mg, Ar, 21 closed-shell molecules: 
H$_2$O, (H$_2$O)$_2$, (H$_2$O)$_3$, (H$_2$O)$_4$, HCN, CH$_4$, C$_2$H$_4$, CO, CO$_2$, H$_2$CO,
NH$_3$, BN, H$_2$O$_2$, C$_6$H$_6$, LiH, NaH, MgH$_2$, AlH, NaOH, MgO, NaF, 
and 4 transition metal atoms and diatomic clusters: Cd, Pd, Cu$_2$, Ag$_2$. 
The aug-cc-pVDZ~\cite{Woon:jcp/100/2975,Dunning:jcp/90/1007,Woon:jcp/98/1358} basis set was used for atoms, 
the TZ(Dunning) basis set was used for molecules except 
for LiH, NaH, MgH$_2$, AlH, NaOH, MgO and NaF where the aug-cc-pVDZ basis 
set was used. For transition-metal containing compounds, a basis set with pseudopotentials 
ecp-sdd-DZ~\cite{Dolg:jcp/86/866,Stoll:jcp/79/5532,Feller:jcc/17/1571,Schuchardt:jcim/47/1045} was employed. 
We studied systems with
pseudopotentials because they are frequently used in solid-state calculations thus giving us an insight into 
behavior of the realistic Green's function in such systems.
Moreover, without using pseudoptentials the grid requirements for such electron rich systems are enormous. 

All systems studied in this work have a small dependence on temperature due to a large 
HOMO-LUMO gap and thus variations of inverse temperature $\beta$ do not change results qualitatively. 
Lowering the temperature (increasing $\beta$) corresponds to decreasing the Matsubara spacing, so that
correspondingly more frequencies are required to reach the same accuracy for frequency sums and energies.
Consequently, to challenge our algorithm we have chosen a relatively large value of inverse temperature 
$\beta=100$ [1/a. u.].

To test the accuracy of our algorithm, we applied it to converged GF2 Green's functions using several 
values of the threshold $\delta = 10^{-n}, n \in \{2,3,4,5,6\}$. For every value of the threshold, we 
used the cubic spline interpolation algorithm to obtain a new small grid --- a set of not necessarily 
equidistant imaginary frequency points satisfying conditions discussed in the previous section. 
A spline evaluated using such a small grid was used 
to calculate the one-body density matrix (eq~\ref{eq:gamma}), total number of electrons, one-body energy 
(eq~\ref{eq:e1b}), and Galitskii--Migdal (eq~\ref{GM_enrg}) and Luttinger--Ward (eq~\ref{LW_enrg}) energies.  

First, we consider the size of new smaller grids used to create a spline and their dependence on the 
value of the threshold $\delta$. The interpolation algorithm applied to a smooth function such as 
Matsubara Green's function produces sets of points with increasing cardinality as the value of 
threshold decreases. This guarantees a monotonic convergence to the original grid (usually containing 
thousands of points) in the limit of $\delta \to 0$, thus making our algorithm controlled and systematically 
convergent when applied to Green's functions. 

The numerical manifestation of the statement above is shown in 
Table~\ref{tab:1}, where we summarized results for several atoms and molecules with different basis sets. 
Before a Green's function calculation is started  the number of points in the full Matsubara grid has to be predetermined. 
This number can be determined by converging the HF energy to a predetermined accuracy by using the HF Green's function. 
The convergence in total HF energy to 10 $\mu$E$_h$ was used to obtain the maximum number of the grid points listed in 
the VIII column of  Table~\ref{tab:1}. Alternatively, one can define an accuracy threshold $\epsilon = \max | A_{ij} |$, 
where $\mathbf{A} = \mathbf{G}^{n}_1 - \mathbf{G}^{a}_1$, is the maximum matrix element of a difference between the 
numerical $\mathbf{G}^{n}_1 = \mathbf{G}(i\omega_n)\cdot i\omega_n$ and the analytical $\mathbf{G}^{a}_1=\mathbf{S}^{-1}$ high-frequency 
coefficient and determine how many points are necessary to converge 
the calculation to match a certain $\epsilon$.
\begin{table*}
\caption{The number of grid points as a function of the threshold $\delta$ used in our cubic spline 
interpolation algorithm for several atoms and molecules\textsuperscript{\emph{a}} as compared to the 
number of points in the input Matsubara grid listed in column VIII. Columns IX$-$X show the number of 
points in the Matsubara grid required to recover the $\mathbf{G}_1$ coefficient of the high-frequency 
expansion of Green's function to $\epsilon=0.1$ and $\epsilon=0.01$ accuracy threshold.}
\label{tab:1}
\begin{ruledtabular}
\begin{tabular}{lcrrrrrrrrr}
\toprule
Atom or         &   Basis      & \multicolumn{5}{c}{$\delta$} & \multicolumn{1}{c}{$N_\omega$} & \multicolumn{2}{c}{$\epsilon$} \\ \cline{3-7} \cline{9-10}
molecule       &      set       &  $10^{-2}$   &    $10^{-3}$  &    $10^{-4}$   &    $10^{-5}$    & $10^{-6}$ & \multicolumn{1}{c}{used}  &  $10^{-1}$        & $10^{-2}$ \\
\hline
Be          &  aug-cc-pVDZ  &  169  & 229 &  337 & 501   & 943   & 3$\cdot 10^{4}$ &  6.7$\cdot 10^{2}$ & 6.8$\cdot 10^{3}$ \\
Mg         &   aug-cc-pVDZ  &  157  & 216 & 328 & 512   &  1027  & 2$\cdot 10^{4}$ & 7.7$\cdot 10^{3}$  & 7.8$\cdot 10^{4}$ \\
NaH       &  aug-cc-pVDZ  &   217  &  343  &  546  &  1046  &  2098  & 2$\cdot 10^{4}$ & 9.1$\cdot 10^{3}$  & 9.1$\cdot 10^{4}$ \\
Ar           &  aug-cc-pVDZ  &  573  & 682 &  841 & 1231 & 1811 &  2$\cdot 10^{5}$ &  1.9$\cdot 10^{4}$  & 1.9$\cdot 10^{5}$\\
NaF       &  aug-cc-pVDZ  &   262   &  417  &   755  &  1502  &  3087  & 2$\cdot 10^{4}$ &   2.7$\cdot 10^{4}$ & 2.7$\cdot 10^{5}$\\
C$_2$H$_4$     &  TZ(Dunning)   &   241  &  384  &  761  &  1931  &  4078   & 2$\cdot 10^{4}$ & 3.1$\cdot 10^{4}$  & 3.1$\cdot 10^{5}$\\
Cd         &  ecp-sdd-DZ    &   255   & 380  &  598  & 1478  & 3196 &  5$\cdot 10^{4}$  & 3.1$\cdot 10^{4}$  & 3.1$\cdot 10^{5}$ \\
MgH$_2$     &  aug-cc-pVDZ &    298  &  469 &  875  &  1693  &  3752  &  3$\cdot 10^{4}$  & 3.2$\cdot 10^{4}$  & 3.2$\cdot 10^{5}$ \\
Ag$_2$\textsuperscript{\emph{b}}    & ecp-sdd-DZ &  294  & 421  & 657  & 1927  &  4064  & 7$\cdot 10^{4}$ & 3.4$\cdot 10^{4}$ & 3.4$\cdot 10^{5}$\\
NH$_3$       &  TZ(Dunning)   &   211  &  344  &  734  &  1612  &  3478  &  2$\cdot 10^{4}$  & 4.6$\cdot 10^{4}$  & 4.6$\cdot 10^{5}$ \\
HCN      &   TZ(Dunning)  &   264  &  411  &  834  &  1941  &   3864  &  2$\cdot 10^{4}$ & 4.7$\cdot 10^{4}$  & 4.7$\cdot 10^{5}$\\
(H$_2$O)$_2$\textsuperscript{\emph{c}}  &   TZ(Dunning)  &   326  &  496  &  885  &  2117  &  4463  & 8$\cdot 10^{4}$   & 6.1$\cdot 10^{4}$  & 6.1$\cdot 10^{5}$\\
(H$_2$O)$_3$\textsuperscript{\emph{c}}  &   TZ(Dunning)  &   252  &  419  &  894  &  2335  &  5074  & 4$\cdot 10^{4}$ &  6.1$\cdot 10^{4}$ & 6.1$\cdot 10^{5}$ \\
H$_2$CO    &  TZ(Dunning)   &   230  &  373  &  770  &  1971  &  4078  &  3$\cdot 10^{4}$ & 6.1$\cdot 10^{4}$  & 6.1$\cdot 10^{5}$\\
C$_6$H$_6$     &   TZ(Eunning) &   322  &  497  &  918  &  1330  &  2581  &  2$\cdot 10^{4}$ & 9.0$\cdot 10^{4}$ & 9.0$\cdot 10^{5}$ \\
\bottomrule
\end{tabular}
\end{ruledtabular}

\textsuperscript{\emph{a}} Experimental geometries were taken from NIST Computational Chemistry Comparison and
Benchmark Database.~\cite{nist}\\
\textsuperscript{\emph{b}} $d$(Ag-Ag)=5.46 a. u.\\
\textsuperscript{\emph{c}} Geometry was taken from ref~\citenum{Wales199865}.\\
\end{table*}

It is clear from Table~\ref{tab:1} that for most systems to recover either the highly accurate HF energy  
or $\mathbf{G}_1$ to at least $\epsilon=0.1$ accuracy frequently more than
10,000 Matsubara frequencies are required. Additionally, we observe that for most systems, besides few atomic examples, 
the size of the grid requiring an $\epsilon$ between $0.1-0.01$ is enough to converge the HF energy to a high accuracy.
Thus, while we can predetermine how large the grid should be, such size of the grids cannot be easily explicitly 
tractable since for small molecules such as those shown in the Table~\ref{tab:1}, if larger basis set are employed, 
the grid can approach a size of 100,000 or more Matsubara frequencies.
This is a numerical explanation why finite temperature Green's function calculations for realistic systems have not yet become routine.  
In this light, our cubic spline interpolation
approximation is an important step towards reliable finite temperature Green's function calculations.
As Table~\ref{tab:1} shows, when using the spline interpolation procedure, the reduction in the size of imaginary
frequency grids is approximately two orders of magnitude if $\delta=10^{-4}$ is used and by about one order for 
lower values of delta. For the most demanding system studied in this work, we only require fewer than 3,000 frequency 
points to produce a new Green's function using spline which is guaranteed to be in a very good agreement 
with the reference one since the threshold value is very small $\delta=10^{-6}$. Overall, when  $\delta=10^{-4}$ is 
used, the number of frequency points necessary to create a spline grid is around 5\% of the original Matsubara grid 
size which is a remarkable reduction. 

To put our efforts in reduction of the Matsubara frequency grid to several thousands points in perspective, it 
is worth mentioning that grids containing the same order of magnitude of 
points are used in quantum chemistry, e.g. in the evaluation of the contribution of the exchange-correlation function in DFT 
and other numerical algorithms~\cite{Becke:jcp/88/2547,Murray:mp/78/997,GILL1993506,Mura:jcp/104/9848,JCC:JCC10211}.

\begin{figure*}[htbp]
\begin{center}
\includegraphics[width=0.98\textwidth]{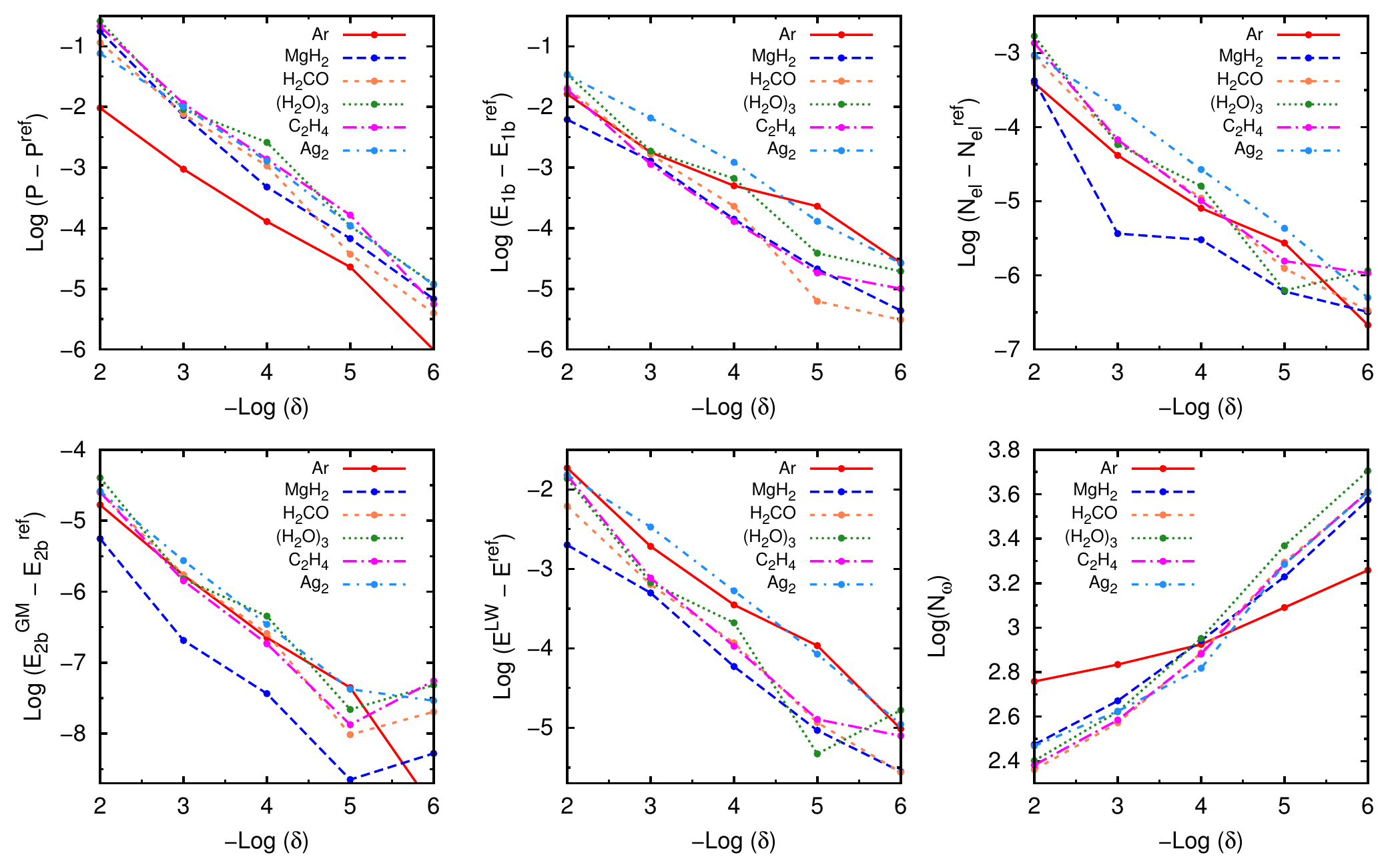}
\caption{Errors in the density matrix ${\rm Log}(\sum_{ij}\left[\mathbf{P} - \mathbf{P}^\text{ref}\right]_{ij})$,
one-body energy ${\rm Log}(E_{1b} -E_{1b}^\text{ref})$, total number of electrons ${\rm Log}(N_{el} -N_{el}^\text{ref})$, 
two-body Galitskii--Migdal energy ${\rm Log}(E^{GM}_{2b} -E_{2b}^\text{ref})$ and 
Luttinger--Ward energy ${\rm Log}(E^{LW} -E^\text{ref})$ as a function of the
threshold ${\rm Log}(\delta)$ for selected atoms and molecules with different basis sets. Last panel
shows the dependence of the grid size ${\rm Log}(N_\omega)$ on the threshold ${\rm Log}(\delta)$.
Log denotes the base 10 logarithm.}
\label{fig:cvg}
\end{center}
\end{figure*}

Next, in Fig.~\ref{fig:cvg}, we examine the convergence of all quantities considered such as the one-particle 
density matrix, one-body energy, etc. as a function of threshold $\delta$. To illustrate the trend, we 
selected a few systems from the test set.

In the first panel of Fig.~\ref{fig:cvg}, we plotted the logarithm of the error in the one-particle 
density matrix defined as $\Delta P = \sum_{ij} \left[ \mathbf{P} - \mathbf{P}^\text{ref}\right]_{ij}$ 
as the function of the threshold $\delta$. For all systems studied in this work, we observed almost perfect 
linear convergence. Thus, an order of magnitude improvement in the accuracy of
the one-particle density matrix can be achieved by decreasing the threshold by a factor of ten. For small 
systems such as atoms, the value of threshold roughly corresponds to the accuracy of the one-particle density
matrix. For bigger systems using the smallest threshold yields inaccurate one-body density matrices. Thus at 
least $\delta=10^{-3}$ should be used if quantities that rely on an accurate determination of a one-particle 
density matrix are of interest.

The one-body energy is calculated using the one-body density matrix. Consequently, we note based on the 
upper center panel of Fig.~\ref{fig:cvg}, that the rate of convergence of the one-body energy is that of the 
one-particle density matrix. The overall accuracy of the one-body energy is worse than that of the 
one-particle density matrix but insignificantly. Nonetheless for all systems considered here the convergence
of about 10 $\mu E_h$ is achieved for $\delta=10^{-6}$.

The total number of electrons is another quantity calculated from the one-particle density matrix.
The $\delta$-dependence of the total number of electrons is shown in the upper right panel of Fig.~\ref{fig:cvg}. 
Overall convergence of the total number of electrons is fast and the accuracy is generally better than one 
of the one-particle density matrix. For the majority of systems considered in this work, it is enough to 
set $\delta=10^{-4}$ to recover the total number of electrons to $10^{-5}$ accuracy. 

The convergence of the Galitskii--Migdal two-body energy is shown in the bottom left panel of Fig.~\ref{fig:cvg}. 
We observe that the Galitskii--Migdal  energy is converging the most rapidly and for
all the systems studied $\delta=10^{-3}$ is enough to achieve a $\mu$E$_h$ accuracy. Thus, the cubic spline 
interpolation algorithm is an extremely efficient way to calculate the two-body energy and to replace a 
simple sum over Matsubara frequencies which is known to be numerically challenging. Small fluctuations in the
Galitskii--Migdal energy observed for $\delta=10^{-5}$ and $\delta=10^{-6}$ are purely numerical artifacts
and only happen after a very good convergence to 0.1 $\mu E_h$ is achieved. 

Next, we examine the convergence of the Luttinger--Ward ($E^\text{LW}$) energy
shown in the bottom center panel of Fig.~\ref{fig:cvg}. The Luttinger--Ward energy converges at a slower rate than
Galitskii--Migdal energy but still an acceptable accuracy of 10 $\mu$E$_h$ can be achieved with $\delta=10^{-5}$ or
$\delta=10^{-6}$ threshold depending on the system under consideration. The oscillations in the Luttinger--Ward
energy, which are numerical artifacts, may also happen but only after initial convergence to less than 10 $\mu E_h$
is achieved.

Finally, last panel of  Fig.~\ref{fig:cvg} shows the sizes of imaginary frequency
grids that correspond to particular value of the threshold $\delta$. As expected, simpler atomic systems 
even with basis sets containing polarization functions generally do not require large grids and changing the 
threshold $\delta$ does not result in a significant change of the size of the imaginary frequency grid, 
indicating that a convergence is achieved with relatively low value of $\delta$. For bigger 
systems with lower spatial symmetry, the convergence of the grid size is slower and larger values of 
the threshold $\delta$ may be necessary.

As the results above indicate, decreasing the threshold $\delta$ indeed results in generating more 
extensive grids with the increase of 1.5$-$2.0 for every order of magnitude decrease of $\delta$ (see 
Table~\ref{tab:1}). This has an important consequence for the computational complexity of realistic 
calculations. In order to achieve an order of magnitude improvement in convergence of the Green's function 
using standard equidistant Matsubara frequency grid the next successive grid must be at least an order of magnitude 
larger than previous one (Fig.~\ref{fig:s_norm}). It follows from Fig.~\ref{fig:cvg} that if a cubic spline 
interpolation algorithm is used, an order of magnitude improvement in the calculated quantity can be 
achieved if the value of threshold $\delta$ is changed by a factor of ten which results in 
changing the number of frequencies necessary only by a factor of 1.5$-$2. This means that upon going to 
bigger systems one should not expect the spline grid to grow as fast as the standard Matsubara frequency grid. 

\begin{figure*}[htbp]
\begin{center}
\includegraphics[width=0.98\textwidth]{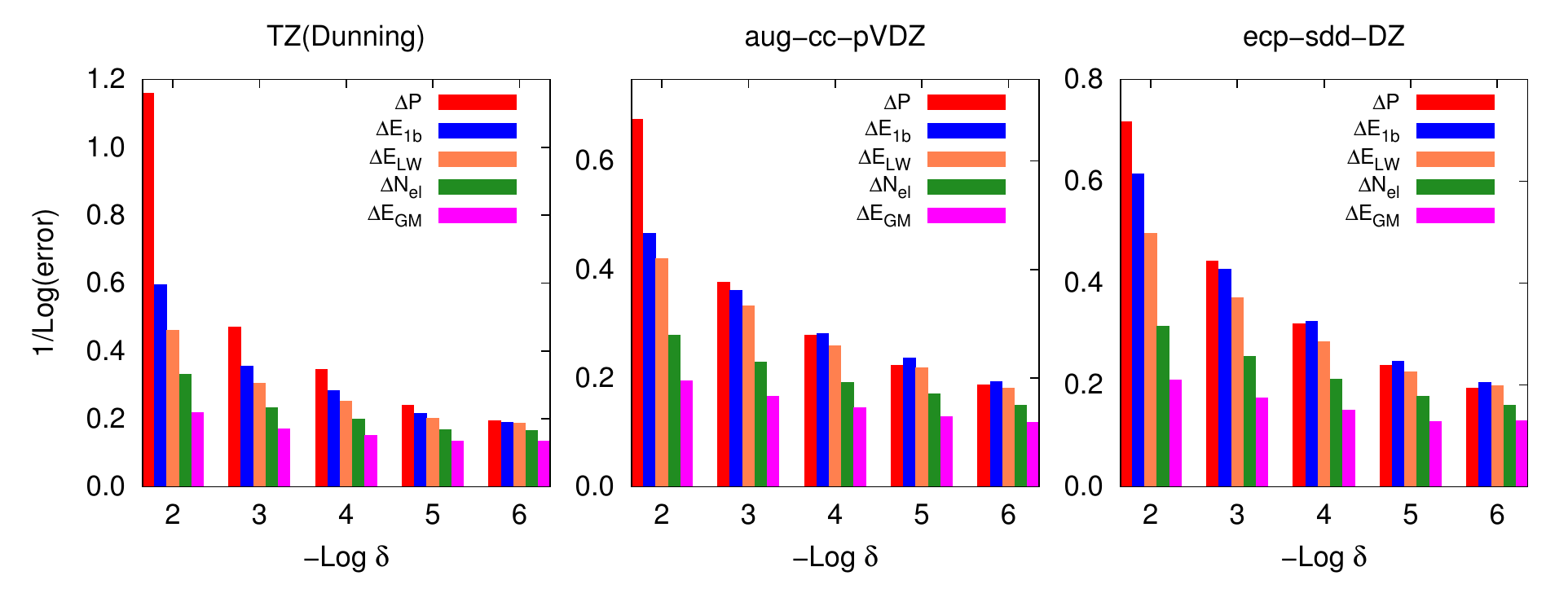}
\caption{The mean average errors in density matrix $\Delta P$, total number of electrons $\Delta N_{el}$, 
one-body energy $\Delta E_{1b}$, two-body energy $\Delta E_{2b}^{GM}$ calculated using Galitskii--Migdal formula
and Luttinger--Ward energy $\Delta E^{LW}$ as a function of the threshold $\delta$
for various basis sets. This composite plot summarizes errors for all the 28 systems considered in this work.
Log denotes the base 10 logarithm.}
\label{fig:bar_all}
\end{center}
\end{figure*}

Finally, since basis set significantly affects the grid requirements, in Fig.~\ref{fig:bar_all}, we plot 
the results for all the 28 systems considered here according to the basis set. We plot 1/Log(error) versus 
$-$Log($\delta$) using bars since we attempt to show errors which differ by orders of magnitude on one plot.
These plots should be read as follows. Each bar represents the mean average error in the calculated quantity 
and longer bars correspond to larger errors.

Fig.~\ref{fig:bar_all} shows the errors in one-particle density matrix, total number of electrons as well as one- 
and two-body energies for three different basis sets employed in our calculations: TZ(Dunning), aug-cc-pVDZ and 
ecp-sdd-DZ. As seen from Fig.~\ref{fig:bar_all}, changing threshold from loose ($10^{-2}$) to tight ($10^{-6}$) 
leads to monotonic improvement in the accuracy for all the quantities considered. For all the basis sets 
employed, the biggest error is in the one-particle density matrix and consequently the one-body energy. 
Smaller errors are observed in the Luttinger--Ward energy and the total number of electrons. 
The error in the two-body energy using the Galitskii--Migdal formula is the smallest and it is almost 
grid size independent. Thus, one can expect that only a small number of grid points can be used to construct 
a spline that is used to evaluate a product of two frequency dependent quantities (like two-body energy from 
eq~\ref{GM_enrg}), while a larger grid is required when the quantity calculated is directly dependent on the 
accuracy of the one-body Green's function (like the one-body density matrix which is related to the Green's 
function by a Fourier transform). By comparing the magnitude of all errors, we conclude that they are the largest 
for the TZ(Dunning) basis set. This is not surprising since we used TZ(Dunning) basis set for bigger molecules 
as opposed to aug-cc-pVDZ basis set used for atoms and smaller molecules. This 
trend is in agreement with regular Matsubara frequency grid requirements shown in Table~\ref{tab:1}.

\section{Conclusions}
\label{conclusions}

If equidistant numerical grids are used, finite temperature Green's function calculations 
of molecular systems or solids in large basis sets seem hardly possible due to highly inefficient
grid spacing. However, since the Matsubara Green's function is a smoothly  and slowly varying function 
of frequency even simple cubic spline interpolation algorithm can help reduce the number 
of grid points at which the Green's function is evaluated explicitly, thereby making realistic 
calculations tractable. We carefully investigated  this idea on a series of atomic and molecular 
calculations with realistic Hamiltonians. We demonstrated that only around 5\% of
the original equidistant Matsubara frequency grid was necessary to obtain very accurate results
for the density matrix or total energy.

Our interpolation algorithm introduces a single threshold parameter---the magnitude of the fourth
derivative, that systematically controls the spline accuracy. To keep the value of this threshold constant 
and below a user desired level, our algorithm detects the regions where the Matsubara Green's function changes rapidly
and ensures that more grid points are used in these regions while fewer points are necessary in the regions with a
slowly changing Green's function.

We established that irrespective of the basis set or the actual system under study, the magnitude of 
the spline fourth derivative is directly proportional to the accuracy of the results. Thus, in a black 
box manner, by changing the value of this parameter we can achieve a desirable high accuracy while 
maintaining a low computational cost. 

One of the most important features of our algorithm shown here is that the growth of the spline grid 
necessary to evaluate the one-body density matrix or energies to a desired accuracy is much slower than 
that of standard equidistant Matsubara frequency grid. While the Matsubara frequency grid grows by an order to magnitude to 
get an order of magnitude improvement in the accuracy, the spline grid only grows by a factor or 1.5$-$2. 
Consequently, to achieve a very high $\mu$E$_h$ convergence of energy with respect to the grid size, 
the number of points at which the Green's function is evaluated explicitly is within only couples of thousands 
while traditional Matsubara frequency grid requires hundreds of thousands of explicit evaluations.


This study, when combined with our recently proposed algorithm for efficient reduction of the size of 
the imaginary time grid in ref~\citenum{Kananenka:jctc/12/564}, is a step towards reliable and computationally 
affordable Green's function calculations  in quantum chemistry and materials science.

\label{conclusions}

\section*{Acknowledgements}

A. A. K., T. N. L, and D. Z. acknowledge support from the U.S. Department of Energy (DOE) (No. ER16391).
A. R. W. acknowledges support from MCubed program at University of Michigan.
E. G. acknowledges support from the Sloan and Simons foundation.


%

\end{document}